\documentclass[twocolumn,showpacs,preprintnumbers,amsmath,amssymb]{revtex4-1}
\usepackage{epsfig}
\usepackage{graphicx}
\usepackage{epsfig}
\usepackage{rotating}
\usepackage{amssymb}
\usepackage{amsmath}
\usepackage{graphicx}

\def\beq{\begin{eqnarray}}
\def\eeq{\end{eqnarray}}

\newcommand{\be}{\begin{equation}}
\newcommand{\ee}{\end{equation}}
\newcommand{\bea}{\begin{eqnarray}}
\newcommand{\eea}{\end{eqnarray}}

\def\bs{\boldsymbol}
\def\bd{\boldsymbol{\mathrm{d}}}
\def\rd{\mathrm{d}}

\newcommand{\ti}{{t_{\mathrm{o}}}}
\newcommand{\tf}{{t_{\mathrm{f}}}}

\newcommand{\av}[1]{\prec #1 \succ}

\begin{document}

\title{Optimal protocols and optimal transport in stochastic thermodynamics}

\author{Erik Aurell${}^{1,2,3}$}
\email{eaurell@kth.se}
\author{Carlos Mej\'ia-Monasterio${}^{4,5}$}
\email{carlos.mejia@upm.es}
\author{Paolo    Muratore-Ginanneschi${}^{5}$}
\email{paolo.muratore-ginanneschi@helsinki.fi}
\affiliation{${}^1$ACCESS Linnaeus Centre, KTH, Stockholm Sweden}
\affiliation{${}^2$Dept.~Computational Biology,  AlbaNova University Centre,
  106 91 Stockholm, Sweden}
\affiliation{${}^3$Aalto    University   School   of    Science,   Helsinki,
  Finland}
 \affiliation{${}^4$University  of Helsinki,  Department of  Mathematics and
   Statistics P.O.  Box 68 FIN-00014, Helsinki, Finland}
\affiliation{${}^5$Laboratory  of Physical  Properties, Department  of Rural
  Engineering, Technical  University of Madrid,  Av.  Complutense s/n,
  28040 Madrid, Spain}

\begin{abstract}
  Thermodynamics  of small systems  has become  an important  field of
  statistical  physics.  They  are  driven out  of  equilibrium by  a
  control, and the question is  naturally posed how such a control can
  be optimized.   We show that  optimization problems in  small system
  thermodynamics are solved  by (deterministic) optimal transport, for
  which very  efficient numerical methods have been  developed, and of
  which  there   are  applications  in   Cosmology,  fluid  mechanics,
  logistics,  and many  other  fields. We  show,  in particular,  that
  minimizing   expected  heat   released   or  work   done  during   a
  non-equilibrium  transition  in finite  time  is  solved by  Burgers
  equation  of Cosmology and  mass transport  by the  Burgers velocity
  field.   Our contribution  hence considerably  extends the  range of
  solvable optimization problems in small system thermodynamics.
\end{abstract}

%05.40.-a: Fluctuation phenomena, random processes, noise, and Brownian motion
%05.40.Jc: Brownian motion 
%05.70.-a: Thermodynamics
%05.70.Ce: Thermodynamic functions and equations of state 
%02.50.Ey: Stochastic processes 
%87.15.H-: Dynamics of biomolecules 
\maketitle

%%%%% -------------------------------------------------------------------------
%%%%% -------------------------------------------------------------------------
%%%%% -------------------------------------------------------------------------
The last two decades has seen a revolution in the understanding of
thermodynamics of small systems driven out of equilibrium. Jarzynski's
equality (JE)~\cite{Jarzynski97} relates an exponential average of the
thermodynamic work $W$ done on a system, driven from an initial
equilibrium state to another final state, to the exponentiated free
energy difference $\Delta F$ between these two states:
\begin{equation} \label{JE} 
\av{e^{-\beta W}} = e^{-\beta\Delta  F} \ .
\end{equation}
Here and in the following $\beta = 1/k_B T$ is the inverse
temperature, $k_B$ the Boltzmann's constant and $\av{\cdot}$ is an
expectation over a non-equilibrium process, specified by a (time- and
state-dependent) driving force or protocol.  JE, and Crook's
theorem~\cite{Crooks00}, from which it follows, has been used to
successfully determine binding free energies of \textit{single}
biomolecules through repeated pulling experiments~\cite{BLR05}, a feat
which had previously been unimaginable.  For stochastic thermodynamics
(the setting of this paper), such transient non-equilibrium
fluctuation relations are comprehensively reviewed in~\cite{ChG07}.  A
counter-part of the transient fluctuation relations are equally
important steady-state fluctuation
relations~\cite{ECM93,ES94,GC95,Kurchan98,LS99,RM-M07,HS07}, but these
fall outside the scope of the present Letter where we consider only
processes in a finite time interval.

The transient non-equilibrium fluctuation relations are
identities; they hold irrespective of the protocol.  Most
quantities of interest however still depend on the protocol, and can
then be varied and optimized. 
A first step in this direction was taken by Schmiedl \& Seifert
 who showed that when pulling
a small system by optical tweezers, (expected) heat released to the 
environment and (expected) work done on the small system
are minimized, not by naively smoothly pulling, but by
protocols with discontinuities \cite{SS07}, a work
which has generated considerable interest in the
field~\cite{G-MSS08,TE08,GD10}.
For technical reasons, the
analysis of Schmiedl \& Seifert was limited to harmonic potentials.

In this Letter we show how such optimization problems in stochastic
thermodynamics (minimizing heat, work, the variance of the JE estimate
of free energy differences) can be mapped to problems of
(deterministic) optimal transport.  The optimal control (for any of
these cases) is determined by the solution of an auxiliary problem.
When optimizing heat or work, this auxiliary problem is none other
than the Burgers equation of fluid dynamics and cosmology, and mass
transport by the Burgers field. Very efficient numerical methods have
been developed to solve such problems, and these methods can be
directly applied.  Our contribution hence extends considerably the
range of solvable optimization problems in stochastic thermodynamics.

%%%%% -------------------------------------------------------------------------
%%%%% -------------------------------------------------------------------------
%%%%% -------------------------------------------------------------------------
{\it Stochastic thermodynamics and optimal protocols:}
We consider dynamics in the overdamped limit described by 
coupled Langevin equations:
\begin{equation}
\label{wd:sde}
\dot{\bs{\xi}_{t}}=-\frac{1}{\tau}\partial_{\boldsymbol{\xi}_{t}}V
\left(\boldsymbol{\xi}_{t},t\right)
+\sqrt{\frac{2}{\tau\beta}}\,\bs{\dot{w}}_{t}\ ,
\end{equation}
with initial value $\bs{\xi}_{\ti}=\bs{x}_{\mathrm{o}}$, drift
$-\partial_{\boldsymbol{\xi}_{t}}V$ and $\dot{\bs{w}}_{t}$ a vector
valued white noise with covariance $\langle \dot{\bs{w}}_{t}
\dot{\bs{w}}_{t'} \rangle = \delta(t-t')$.  The mobility is
$\tau^{-1}$ and $\beta$ the inverse temperature.  For times $t<\ti$
the potential is $V(\boldsymbol{x},t)=U_{\mathrm{o}}(\boldsymbol{x})$,
and for times $t>\tf$ is $V(\boldsymbol{x},t) =
U_{\mathrm{f}}(\boldsymbol{x})$.  In the control interval $[\ti,\tf ]$
we allow the potential to be an explicit function of time
$V(\boldsymbol{x},t)=U(\boldsymbol{x},t)$ \emph{eventually}
discontinuous at the boundaries $V(\boldsymbol{x},\ti) =
\wp_{\mathrm{o}}\,U_{\mathrm{o}}(\boldsymbol{x}) +
(1-\wp_{\mathrm{o}})\,U(\boldsymbol{x},\ti)$ and
$V(\boldsymbol{x},\tf) = \wp_{\mathrm{f}}\,U(\boldsymbol{x},\tf) +
(1-\wp_{\mathrm{f}})\,U_{\mathrm{f}}(\boldsymbol{x})$ with
$0\leq\wp_{\mathrm{i}}\leq 1$, $\mathrm{i} =
\{\mathrm{o},\mathrm{f}\}$.
For single stochastic trajectories we define $\delta \mathcal{W}$, the
Jarzynski work \cite{Jarzynski97}, and $\delta Q$, the heat released
into the heat bath, in as~\cite{Sekimoto98}
\begin{eqnarray}
\label{def:work}
&&\delta W = \int_{\ti}^{\tf}\partial_{t}V(\boldsymbol{\xi}_{t},t)~\rd t \ ,
\\
\label{def:heat}
&&\delta Q =-\int_{\ti}^\tf \! \dot{\boldsymbol{\xi}_{t}}
\cdot\partial_{\boldsymbol{\xi}_{t}}V(\boldsymbol{\xi}_{t},t) ~\rd t \ .
\end{eqnarray}
The difference $\delta W-\delta Q$ satisfies $1^\mathrm{st}$ law,
\emph{i.e.} is the integral of an exact differential, if and only if
the stochastic integral in $\delta Q$ is defined in the sense of
Stratonovich.
The Stratonovich integral is the limit of Riemann sums where products
$V(\boldsymbol{\xi}_{t},t)dw_{t}$ are discretized according to to the
\emph{mid-point} prescription \emph{i.e.}
$V((\boldsymbol{\xi}_{t_{i}}+\boldsymbol{\xi}_{t_{i+1}})/2,
\bar{t}_{i})(w_{t_{i+1}}-w_{t_{i}})$ for $t\in [t_{i},t_{i+1}]$ and
$\bar{t}_{i}$ an arbitrary interpolation rule for $t$. Thus, the
expression of the first law over $[\ti,\tf]$
\begin{eqnarray}
\label{def:DV}
\delta W-\delta Q=V(\boldsymbol{\xi}_{\tf},\tf)-V(\boldsymbol{\xi}_{\ti},\ti) \ .
\end{eqnarray}
does not require $\wp_{\mathrm{f}}=\wp_{\mathrm{o}}=1/2$ for
discontinuities in the time argument although the choice may appear
otherwise appealing.

The stochastic differential equations (\ref{wd:sde}) lead to a
(control-dependent) probability density $m(\boldsymbol{x},t)$ evolving
according the Fokker-Planck equation
and the expectation value of a local quantity $\mathcal{G}$ is
\begin{equation}
\label{uncondE}
\prec\,\mathcal{G}\left(\bs{\xi}_{t},t\right)\,\succ=
\int \bd x~ m(\boldsymbol{x},t)
\mathcal{G}\left(\bs{x},t\right) \ .
\end{equation}
Straightforward application of It\^o lemma (see e.g.\cite{Durrett})
yields 
\begin{equation} \label{heat:heat}
\av{\delta{Q}}= -\int_{\ti}^\tf\rd t\,%\av{ 
\prec\,\mathfrak{L}_{\boldsymbol{\xi}_{t}}^{[-\partial_{\boldsymbol{\xi}_{t}}U]} U\,\succ \ ,
\end{equation}
for $\mathfrak{L}_{\boldsymbol{x}}^{[\boldsymbol{b}]} :=
\frac{\boldsymbol{b}}{\tau} \cdot \partial_{\boldsymbol{x}} +
\frac{1}{\beta\,\tau}\partial_{\boldsymbol{x}}^{2}$ the generator of
the diffusion process with drift $\boldsymbol{b}$.  Given initial and
final states, the minimal {\it variance} of the heat (or work) can be
written as a Kullback-Leibler distance between a controlled and
uncontrolled process, and this connection has been thoroughly explored
in the literature~\cite{DaiPra91,FH05}.  We will here be concerned
with $\av{\delta{Q}}$, $\av{\delta{W}}$ and exponentially weighted
functionals of the heat or the work.

%%%%% -------------------------------------------------------------------------
%%%%% -------------------------------------------------------------------------
%%%%% -------------------------------------------------------------------------
{\it Burgers equation in optimal stochastic control:} We first focus
on heat minimization.  Following \cite{GM83}, we look for a function
$A(\boldsymbol{x},t)$ such that when evaluated along
$\boldsymbol{\xi}_{t}$
\begin{eqnarray}
\label{vp:principle}
0=\int_{\ti}^{\tf}dt \prec\,\partial_{t}A+
\mathfrak{L}_{\boldsymbol{\xi}_{t}}^{[-\partial_{\boldsymbol{\xi}_{t}}U]}(A+U)\,\succ \ .
\end{eqnarray}
If such function can be found, the identity
\begin{eqnarray}
\label{differential}
\prec\,\delta \mathcal{Q}\,\succ=
\prec\,A\left(\boldsymbol{\xi}_{\ti},\ti\right)-
A\left(\boldsymbol{\xi}_{\tf},\tf\right)\,\succ \ ,
\end{eqnarray}
holds true as $\prec\,(\partial_{t} +
\mathfrak{L}_{\boldsymbol{\xi}_{t}}^{[-\partial_{\boldsymbol{\xi}_{t}}U]})A\,\succ$
is the average of an exact stochastic differential.
A sufficient condition for (\ref{vp:principle}) to be satisfied is the
so-called dynamic programming equation (DPE) $\partial_{t}A +
\mathfrak{L}_{\boldsymbol{x}}^{[-\partial_{\boldsymbol{x}}U]}(A+U) =0$
which for any given value of $U$ yields a linear, backwards in time
evolution for $A$.  The stationarity condition for DPE is obtained by
taking the functional variation of (\ref{vp:principle}) with respect
to $U$.
Introduce the (fictitious) potential $R(\boldsymbol{x},t)$
corresponding to the state $m(\boldsymbol{x},t)$ if it would have been
in equilibrium \textit{i.e.}  $R=\frac{1}{\beta}\log m$.  Then the
variation of $U$ yields the condition
\begin{equation} \label{minQ:stat}
\mathfrak{L}_{\boldsymbol{x}}^{[\partial_{\boldsymbol{x}}R]}(A-2\,U-R)=0 \ ,
\end{equation}
which is satisfied independently of $\partial_{\boldsymbol{x}}R$ if
the potential is
\begin{eqnarray}
\label{optimum}
U_{*}=\frac{A-R}{2}+\phi \ ,
\end{eqnarray} 
where $\phi$ is an arbitrary function of time alone.
The optimal control potential is therefore the solution of the coupled
\textit{backwards}- \textit{forwards} equations
\begin{eqnarray}
\label{eq:HBJ}
&&\partial_{t}A + 
\mathfrak{L}_{\boldsymbol{x}}^{[\partial_{\boldsymbol{x}}\frac{R-A}{2}]}\frac{A+R}{2}=0 \ ,
\\
\label{eq:Fokker-Planck}
&&\partial_{t} m + \partial_{\bs{x}}\cdot [\frac{\partial_{\bs{x}} 
(R-A)}{2\tau}m] = \frac{1}{\beta\tau}\partial_{\bs{x}}^2 m \ .
\end{eqnarray}
respectively obtained by plugging (\ref{optimum}) into the DPE and
Fokker-Planck equations.  We note that the Fokker-Planck equation has
the property that if we split the drift into an equilibrium piece
$\partial_{\boldsymbol{x}}R$ and a remainder specified by the gradient
of
\begin{equation}\label{eq:drift-split}
\psi =-\frac{A+R}{2} \ ,
\end{equation}
then it becomes the deterministic transport equation in the gradient
of the remainder:
\begin{equation}\label{minQ:transport}
\partial_{t}m + \frac{1}{\tau}\partial_{\bs{x}}\cdot 
[\left(\partial_{\bs{x}}\psi\right)m] = 0 \ .
\end{equation}
It is a perhaps surprising fact that using the definitions of $R$ and
$\psi$ and the Fokker-Planck equation (\ref{eq:Fokker-Planck}) reduces
(\ref{eq:HBJ}) to simply
\begin{equation}\label{minQ:heat}
\partial_{t}\psi+\frac{\parallel\partial_{\bs{x}}
\psi\parallel^{2}}{2\tau} = 0 \ .
\end{equation}
Equation (\ref{minQ:heat}) is Burgers equation (for the velocity
potential), and equation (\ref{minQ:transport}) is the equation of
mass transport by the corresponding velocity field. These two equation
are the first main result of this paper: we have reduced a complicated
stochastic optimization problem to a classical problem of optimal
deterministic transport. In addition, contrasting (\ref{vp:principle})
with the expression of the work imposed by the first law, it is
readily seen that \emph{work} optimization brings about the same
evolution equations (\ref{minQ:transport}), (\ref{minQ:heat}) now
complemented by the \emph{final} boundary condition
\begin{eqnarray}
\label{minW:bc}
A(\boldsymbol{x},\tf)=V(\boldsymbol{x},\tf)=
-[R(\boldsymbol{x},\tf)+2\,\psi(\boldsymbol{x},\tf)] \ .
\end{eqnarray}
It is worthwhile remarking that the occurrence of final time
constraints is a consequence of the backwards time evolution of the
DPE and is a general feature of variational principles in the presence
of boundary cost terms \cite{GM83}.  We now discuss how this transport
problem can be solved, first if the initial and final states are
given, and then if the initial state and the final control are given.

\textit{Optimal heat between given initial and final states:} The
meaning of Burgers equation in~(\ref{minQ:heat}) is somewhat peculiar
in that it arises from a mixed forwards-backwards problem. In other
words, it is not reasonable to regularize (possible) shocks (in the
future or in the past) by either adding $+\nu\partial^2_{\bs{x}}\psi$
or $-\nu\partial^2_{\bs{x}}\psi$ on the right hand side;
equation~(\ref{minQ:heat}) should make sense in both directions.  On
the other hand, without shocks the solutions of Burgers equation are
free-streaming motion, which we can specify by a \textit{inverse
  Lagrangean map} $\bs{x}_{\mathrm{o}} = \bs{x}_{\mathrm{f}} -
(\tf-\ti) \bs{v}(\boldsymbol{x}_{\mathrm{f}},\tf)$
where the velocity (constant along streamlines) is
$\bs{v}(\boldsymbol{x}_{\mathrm{f}},\tf)=
\frac{1}{\tau}\partial_{\bs{x}_{\mathrm{f}}}\psi(\boldsymbol{x}_{\mathrm{f}},\tf)$.
By mass conservation the inverse Lagrangean map must satisfy the
Monge-Amp\`ere equation
\begin{equation}
\label{eq:monge-ampere}
\| \det \frac{\partial x_o}{\partial x_{\mathrm{f}}} \| = 
\frac{m_{\mathrm{f}}(\boldsymbol{x}_{\mathrm{f}})}{m_o(\boldsymbol{x}_o)} \ ,
\end{equation}
where $m_{o}(\boldsymbol{x})\equiv m(\boldsymbol{x},\ti)$ is the
initial state and $m_{\mathrm{f}}(\boldsymbol{x})\equiv
m(\boldsymbol{x},\tf)$ is the final state.  In 1D this equation is
immediately solved in terms of the cumulative mass functions
$\frac{dM_{\mathrm{f}}}{dx}=m_{\mathrm{f}}$ and
$\frac{dM_o}{dx}=m_o$. The inverse Lagrangean map is then determined
by
$M_o(\boldsymbol{x}_o)=M_{\mathrm{f}}(\boldsymbol{x}_{\mathrm{f}})$. For
higher dimensions we note that for free-streaming motion
\begin{eqnarray}
\label{lagmap}
\hspace{-0.2cm}
\bs{x}_{\mathrm{o}}=\partial_{\boldsymbol{x}}
\left[\frac{\parallel\bs{x}_{\mathrm{f}}\parallel^2}{2}-
\frac{\tf\!-\!\ti}{\tau}\psi(\boldsymbol{x}_{\mathrm{f}},\tf)\right]
:=\boldsymbol{\Psi}(\boldsymbol{x}_{\mathrm{f}};\tf,\ti) \ ,
\end{eqnarray}
and (\ref{eq:monge-ampere}) becomes a partial differential equation in
a scalar field $\Psi$:
\begin{equation}
\label{eq:monge-ampere-reduced}
\| \det \frac{\partial^2 \Psi}{\partial x_{\mathrm{f}}^{\alpha}
\partial x_{\mathrm{f}}^{\beta}} \| = \frac{m_{\mathrm{f}}(\bs{x}_{\mathrm{f}})}
{m_{\mathrm{o}}(\partial_{\bs{x}_{\mathrm{f}}}\Psi)} \ .
\end{equation}
Combining (\ref{differential}) with (\ref{eq:drift-split}) the optimal
released heat can be written as
\begin{eqnarray}
\label{eq:optimal-released-heat-o}
\av{\delta Q} =
 -\frac{1}{\beta}\Delta S +2\prec\,\psi(\boldsymbol{\xi}_{\tf},\tf)
-\psi(\boldsymbol{\xi}_{\ti},\ti)\,\succ \ ,
\end{eqnarray}
where $\Delta S=-\beta\prec\,\ln
m_{\mathrm{f}}(\boldsymbol{\xi}_{\tf}) -\ln
m_{\mathrm{o}}(\boldsymbol{\xi}_{\ti})\,\succ$ is the entropy
change. Similarly, the minimal expected work is
\begin{equation}
\label{eq:optimal-work}
\av{\delta W} = \prec\,V(\boldsymbol{\xi}_{\tf},\tf)-
V(\boldsymbol{\xi}_{\ti},\ti)\,\succ + \av{\delta Q} \ ,
\end{equation}
provided $R$ and $\psi$ satisfy (\ref{minW:bc}). In both cases the
difference $2\prec\,\psi(\boldsymbol{\xi}_{\tf},\tf) -
\psi(\boldsymbol{\xi}_{\ti},\ti)\,\succ$ which represents dissipated
work, can also be written
\begin{equation}\label{eq:W-diss}
W_{diss}=\av{\frac{\|\bs{\xi}_{\tf}-\boldsymbol{\Psi}
(\bs{\xi}_{\tf};\tf,\ti)\|^2\tau}{\tf-\ti}} \ .
\end{equation}
Equation~(\ref{eq:W-diss}) means that the initial and final states can
be specified by mass points $\{x_0^{(1)},x_0^{(2)},\ldots,x_0^{(N)}\}$
and $\{x_f^{(1)},x_f^{(2)},\ldots,x_f^{(N)}\}$, and a possible inverse
Lagrangean map by a one-to-one assignment $x_f^{(i)}\to x_0^{(j)}$.
The inverse Lagrangean map solving (\ref{eq:monge-ampere-reduced}) is
then given by the assignment which minimizes the quadratic cost
function~(\ref{eq:W-diss}), an approach which has been used with great
success to reconstruct velocity fields in the early
universe~\cite{MoFrMaSo2003,MoTuFr2007}.  The interpretation of this
quadratic cost function as dissipated work in stochastic
thermodynamics is, up to our knowledge, new.

\textit{Optimal heat with given final control:} A setting which is
closer to the problem of minimizing work discussed by Schmiedl and
Seifert~\cite{SS07} is when the final control $V(\boldsymbol{x},\tf)$
is specified, but not the final state.  Let hence the initial
potential be
$U_{\mathrm{o}}(\boldsymbol{x})=\parallel\boldsymbol{x}\parallel^{2}/2
+\bar{U}_{\mathrm{o}}$ and the initial state be
$m_{\mathrm{o}}(\boldsymbol{x})\sim \exp\{-\beta\,
U_{\mathrm{o}}(\boldsymbol{x})\}$ and the final potential be
$U_{\mathrm{f}}(\boldsymbol{x}) = c\,\parallel\boldsymbol{x} -
\boldsymbol{h}\parallel^{2}/2+\bar{U}_{\mathrm{f}}$ with $c>0$ and
$\bar{U}_{\mathrm{o}}\,,\bar{U}_{\mathrm{f}}$, arbitrary constants.
By (\ref{optimum}) and (\ref{eq:drift-split}) $U_{*}$ satisfies
\begin{eqnarray}
\label{optimum_psi}
U_{*}(\boldsymbol{x},t)=-[\psi(\boldsymbol{x},t)+R(\boldsymbol{x},t)]
+\phi(t) \ .
\end{eqnarray}
The function $\phi$ can be, however, set to zero as the heat depends
only upon the spatial gradient of $U_{*}$. Since for the heat there
are no further conditions on $\psi$ and $R$, we can set
$U_{*}(\boldsymbol{x},\tf) = U_{\mathrm{f}}(\boldsymbol{x}) =
V(\boldsymbol{x},\tf)$. The problem can be then solved by a Gaussian
Ansatz for the measure i.e.
\begin{eqnarray}
\label{Gauss_Ansatz}
R(\boldsymbol{x},t)=-\frac{\parallel\boldsymbol{x}-
\boldsymbol{\mu}_{t}\parallel^{2}}
{2\,\sigma_{t}^{2}}+\frac{d}{2\,\beta}\ln\frac{1}{\sigma_{t}^{2}} \ .
\end{eqnarray}
We can then use (\ref{optimum_psi}) to write
$\psi(\boldsymbol{x},\tf)$ in terms of (\ref{Gauss_Ansatz}) and
$U_{\mathrm{f}}$ hence obtaining $\boldsymbol{\Psi}$ by
(\ref{lagmap}).  Then, plugging $\boldsymbol{\Psi}$ into
(\ref{eq:monge-ampere-reduced}) yields
\begin{eqnarray}
\label{eq:heat_Gauss}
\boldsymbol{\mu}_{\tf}=\frac{c\,T\,\boldsymbol{h}}{Tc\!+\!\tau} 
\hspace{0.25cm}\&\hspace{0.25cm}
\sigma_{\tf} =\frac{2\,T}{\sqrt{4T(cT\!+\!\tau)\!+\!\tau^{2}}\!-\!\tau} \ ,
\end{eqnarray}
for $T:=\tf-\ti$ and after straightforward algebra
\begin{eqnarray}
\label{time_dep}
\boldsymbol{\mu}_{t}=\frac{t-\ti}{T}\boldsymbol{\mu}_{\tf}\hspace{0.25cm}
\&\hspace{0.2cm}\sigma_{t}=1+\frac{t-\ti}{T}(\sigma_{\tf}-1) \ ,
\end{eqnarray}
for any $t\in [\ti,\tf]$.
Finally the optimal heat and drift are
\begin{eqnarray}
\label{eq:optimal-released-heat}
&&\hspace{-0.4cm}\av{\delta Q}=
\frac{\tau\,\parallel\boldsymbol{\mu}_{\tf}\parallel^{2}}{T}
+\frac{d}{\beta}\left[\ln \frac{1}{\sigma_{\tf}}+
\frac{\tau\,(\sigma_{\tf}-1)^{2}}{T}\right] \ ,
\\
\label{eq:optimal-released-heat_drift}
&&\hspace{-0.3cm}-\,\partial_{\boldsymbol{x}}U_{*}=\frac{\boldsymbol{\mu}_{t}
-\boldsymbol{x}}{\sigma_{t}^{2}} + \frac{\tau\,[\boldsymbol{x}\,
(\sigma_{\tf}-1)+\boldsymbol{\mu}_{\tf}]}{T\,\sigma_{t}} \ .
\end{eqnarray}
As expected the results do not depend on
$\bar{U}_{\mathrm{o}}\,,\bar{U}_{\mathrm{f}}$.  Whilst the state
density (\ref{Gauss_Ansatz}) is continuous for all $t\in[\ti,\tf]$,
the optimal drift (\ref{eq:optimal-released-heat_drift}) exhibits a
discontinuity at $t=\ti$ as discussed in \cite{G-MSS08}.

\emph{Optimal work with given final control:}
Work optimization, as considered in \cite{SS07}, exhibits more subtle
features. The final condition (\ref{minW:bc}) together with
(\ref{optimum_psi}) now yield
\begin{eqnarray}
\psi(\boldsymbol{x},t_{f})=-\frac{(1-\wp_{\mathrm{f}})\,
[U_{\mathrm{f}}(\boldsymbol{x})+R(\boldsymbol{x},\tf)]
}{2-\wp_{\mathrm{f}}}+\phi(\tf) \ .
\end{eqnarray}
Using the Gaussian Ansatz (\ref{Gauss_Ansatz}) and proceeding as for
the heat we find that (\ref{time_dep}),
(\ref{eq:optimal-released-heat_drift}) still hold true but the final
mean and variance are now given by
\begin{eqnarray}
\boldsymbol{\mu}_{\tf}=\frac{c\,T\,\boldsymbol{h}\,\tilde{\wp}_{\mathrm{f}}}
{\tilde{\wp}_{\mathrm{f}}\,T\,c+2\,\tau} 
\hspace{0.3cm}\&\hspace{0.3cm}
\sigma_{\tf} =\frac{\tilde{\wp}_{\mathrm{f}}\,T}
{K} \ ,
\end{eqnarray}
with $K = \sqrt{\tilde{\wp}_{\mathrm{f}}\,T\,
  (\tilde{\wp}_{\mathrm{f}}\,c\,T + 2\,\tau) + \tau^{2}} -\tau$, and
$\tilde{\wp}_{\mathrm{i}}:=(1-\wp_{\mathrm{i}})/(1-\wp_{\mathrm{i}}/2)$,
$\mathrm{i}=\{\mathrm{o},\mathrm{f}\}$. As before, drift and density
do not depend upon $\phi$ nor $\bar{U}_{\mathrm{o}}$,
$\bar{U}_{\mathrm{f}}$.  They, however, depend upon the shape of the
discontinuities of the control $V$ at the boundary.  Note that for any
$c\,>\,1$, $\sigma_{\tf}$ is a decreasing function of $T$ such that
$1\geq\,\sigma_{\tf}\geq 1/\sqrt{c}$.  The corresponding expression of
the optimal work is
\begin{eqnarray}
\label{}
\lefteqn{\hspace{-0.4cm}
\prec\,\delta \mathcal{W}\,\succ=
\frac{4-3\,\tilde{\wp}_{\mathrm{o}}}{4\,(2-\tilde{\wp}_{\mathrm{o}})}\left\{
\frac{d \,\tilde{\wp}_{\mathrm{f}}}{\beta}\ln\frac{1}{\sigma_{\tf}}
+\frac{2\,\tau\,d\,(1-\sigma_{\tf})}{\beta\,T}\right.} 
\nonumber\\&&
\left.\hspace{-0.4cm}
+\frac{2\,\tau \parallel\boldsymbol{\mu}_{\tf}\parallel^{2}}{T}
\frac{T\,\tilde{\wp}_{\mathrm{f}}+2\,\sigma_{\tf}\,\tau}
{T\,\tilde{\wp}_{\mathrm{f}}+2\,(1-\sigma_{\tf})\,\sigma_{\tf}\,\tau}\right\}
+\Delta\bar{U} \ ,
\end{eqnarray}
where $\Delta \bar{U} = (1-\wp_{\mathrm{f}})\bar{U}_{\mathrm{f}} +
\wp_{\mathrm{f}}\,\phi(\tf) - (1-\wp_{\mathrm{o}})\,\phi(\ti) -
\wp_{\mathrm{o}}\bar{U}_{\mathrm{o}}$ can always be set to zero
exploiting the arbitrariness of the function $\phi(t)$.  It is
straightforward to verify that the examples considered in \cite{SS07}
are worked out for the case
$(\wp_{\mathrm{o}},\wp_{\mathrm{f}})=(1,0)$ and that as such they are
a special case of the formulas given above.

%%%%% -------------------------------------------------------------------------
%%%%% -------------------------------------------------------------------------
%%%%% -------------------------------------------------------------------------
{\it Optimizing the variance of the Jarzynski estimator:} We now turn
our attention to a different expectation value.  The Jarzynski
Equality (JE) is an equality in expectation~(\ref{JE}), but does not
hold for a finite number of samples \cite{GRB03}.  Let there be $N$
independent measurements of the work; then the free energy difference
is estimated as $\Delta F = -\beta^{-1}\ln (\frac{1}{N}
\sum_{i=1}^N e^{-\beta W_i})$, 
with a statistical error determined by $\hbox{Var}\left[ e^{-\beta
    W}\right]/N$.  Moreover, expectation and variance of a finite
sampling will depend upon the details of the drift.
It therefore makes sense to study the expectation value
$g_{\lambda}(\boldsymbol{x},t) = \av{e^{-\lambda\beta
    W}}_{\boldsymbol{x},t}$,
where we understand that the noise in the stochastic differential
equation (\ref{wd:sde}) is at inverse temperature $\beta$, and that
the initial state is in equilibrium at the same temperature.
Using the approach of~\cite{LS99} $g_{\lambda}$ can be shown to
satisfy for any given $U$ a controlled diffusion
equation
which we can write for $A_{\lambda}=-\frac{1}{\lambda\,\beta}\log
g_{\lambda}$ (note that $g_{0}=1$ by definition) as
\begin{eqnarray}\label{eq:controlled-second-exponential-moment} 
\hspace{-0.3cm}
\mathfrak{L}_{\boldsymbol{x}}^{[(2\,\lambda-1)\partial_{\boldsymbol{x}}U]}A_{\lambda}  =
-\mathfrak{L}_{\boldsymbol{x}}^{[(\lambda-1)\partial_{\boldsymbol{x}}U]}U
+\frac{\lambda}{\tau}\parallel\partial_{\boldsymbol{x}}A_{\lambda}\parallel^{2} \ .
\end{eqnarray} 
The extremum condition for the drift gives
\begin{equation}
\partial_{\boldsymbol{x}}U_{*}=\partial_{\bs{x}}
\frac{(1-2\,\lambda)\,A_{\lambda}-R}{2\,(1-\lambda)} \ .
\end{equation}
If we again, as in (\ref{eq:drift-split}), split the drift into an
equilibrium piece and a remainder
\begin{equation}
\partial_{\boldsymbol{x}}\psi_{\lambda}=-(1-2\,\lambda)\partial_{\boldsymbol{x}}
\frac{A_{\lambda}+R}{2\,(1-\lambda)} \ ,
\end{equation}
we obtain the generalized optimal transport equations
\begin{eqnarray}
\label{eq:lamba-mass}
&& \hspace{-0.9cm}\partial_{t}m
+\frac{1}{\tau}\partial_{\boldsymbol{x}}\cdot\left(m\,
\partial_{\boldsymbol{x}}\psi_{\lambda}\right)=0 \ ,
\\
\label{eq:lamba-mom}
&&\hspace{-0.9cm}\partial_{t}\psi+\frac{\parallel\partial_{\boldsymbol{x}}
\psi\parallel^{2}}{2\tau(1\!-\!2\lambda)}
+\frac{\lambda\,(\partial_{\boldsymbol{x}}^{2}\psi)}{\beta\,\tau\,(1\!-\!\lambda)}
=\frac{\lambda\,(\partial_{\boldsymbol{x}}\psi)\!\cdot\!(
\partial_{\boldsymbol{x}}m)}{m\,\beta\,\tau\,(\lambda\!-\!1)} \ .
\end{eqnarray}
These equations are not immediately solved, and deserve further study.

%%%%% -------------------------------------------------------------------------
%%%%% -------------------------------------------------------------------------
%%%%% -------------------------------------------------------------------------
In summary, we have shown how stochastic optimization problems are
solved by the methods of optimal control. The solution is built on an
auxiliary problem of optimal transport. When minimizing heat or work
of a small system this optimal transport is a classic of fluid
mechanics and cosmology, namely Burgers equation.  Between any
prescribed initial and final states, these problems can be solved
numerically with the Monge-Amp\`ere-Kantorovich method, introduced to
reconstruct velocity fields in the early Universe.  Boundary cost
contributions to the work, penalizing discontinuous controls and hence
overcoming ambiguities in the definition of the free energy, can be
easily handled in the formalism in the form of Lagrange multipliers,
and solved by fast-Legendre transforms methods for, at least, any
convex potential.  The direct connection between optimal transport and
optimal protocols in small system thermodynamics was wholly
unexpected, and is promising, as it applies to a whole wide class of
related optimization problems.

%%%%% -------------------------------------------------------------------------
%%%%% -------------------------------------------------------------------------
%%%%% -------------------------------------------------------------------------
%\textbf{\emph{Acknowledgments}}:
This work was supported by  the Swedish Research Council (E.A) through
Linnaeus  Center ACCESS and  Academy of  Finland center  of excellence
``Analysis and Dynamics Research'', and  Academy of Finland as part of
it   Distinguished   Professor   program   grant   129024.    C.~M.-M.
acknowledges  support  from  the  European Research  Council  and  the
Academy of Finland.

\bibliography{optimal}% 
\end{document}